# Study of electromagnetic wave propagation in active medium and the equivalent Schrödinger equation with energy-dependent complex potential

H. Bahlouli[+], A. D. Alhaidari, and A. Al Zahrani
*Physics Department, King Fahd University of Petroleum & Minerals, Dhahran 31261, Saudi Arabia*
*and*
E. N. Economou
*University of Crete, 711 10 Heraklio,Crete, Creece*
*And*
*IESL, FORTH, P. O. Box 1527, Heraklio, Crete, Greece*

We study the massless limit of the Klein-Gordon (K-G) equation in 1+1 dimensions with static complex potentials as an attempt to give an alternative, but equivalent, representation of plane electromagnetic (em) wave propagation in active medium. In the case of dispersionless em medium, the analogy dictates that the potential in the K-G equation is complex and energy-dependent. In the non-relativistic domain we study an analogous inertial system by considering wave packet propagation through a complex potential barrier and solve the time-independent Schrödinger equation with a potential that has the same energy dependence as that of the K-G equation. The behavior of these solutions is compared with those found elsewhere in the literature for the propagation of electromagnetic plane waves in a uniform active medium with complex dielectric constant. Our study concluded that the discrepancy between the time dependent and stationary results is due to the energy poles crossing the real axis in the complex energy plane. It was demonstrated unambiguously that there is a frequency ( energy ) and size dependent gain threshold above which the stationary results become unstable. This threshold corresponds to the value of the gain at which the pole crosses the real axis.

## I. INTRODUCTION

The interest in amplification effects of classical and quantum waves in disordered media has been strongly motivated by recent experimental results on amplification of light [ 1 ]. The amplification was shown to strongly enhance the coherent back-scattering and consequently increase reflection. These results on the reflection naturally lead one to expect an enhancement of the transmission in such amplifying systems. However, for amplifying periodic systems, many workers [ 2 ] found that the transmission coefficient starts increasing exponentially up to a certain maximum where it oscillates and then decreases exponentially. Actually it was generally shown by Beenakker et al. [ 2 ] that there is a dual symmetry between absorption and amplification for the propagation of radiation through a disordered medium with a complex dielectric constant. Thus the above work confirmed the paradoxial result that stimulated emission of radiation suppresses the transmission through the system. The optical transmission through a segment of complex dielectric material [ 3 ] or the analogous electronic transmission through a complex scattering potential [ 4 ] exhibited a transition from amplification to absorption at a critical value of the gain ( or length ) when treated using the stationary wave equation or the stationary Schrödinger equation. However, when the time

[+] E-mail: bahlouli@kfupm.edu.sa (corresponding author)

dependent wave equation or Schrödinger equations were solved for an initial pulse by the finite difference time domain ( FDTD ) method no region of absorption was observed [ 5 ] even for values of the gain above threshold. Thus it seems that for systems with gain the stationary solutions become irresponsive to time dependent perturbation for gains above a certain threshold. Soukoulis et al. [ 4-5 ] correctly pointed out the nature of the discrepancy between the time dependent wave equation and the stationary one. However, there is no satisfactory explanation of the origin of this apparent paradox.

It is the purpose of this work to try to clarify the origin of the discrepancy between the time dependent and stationary wave equations in the case of systems with gain. Since the above mentioned problem seems to be common to both wave and Schrödinger evolution equations with gain we try to keep this parallelism in our analysis setting aside the issue of practical relevance for electronic systems. In optical systems one can phenomenologically understand the increase of light intensity due to an increase in photons by means of coherent amplification, as by stimulated emission of radiation in an active lasing medium. However, in electronic systems and due to particle number conservation one cannot imagine such a violation of current conservation. To establish the connection between the electromagnetic and inertial systems we start with the relativistic massless Klein-Gordon (K-G) equation which should reproduce the wave equation. From there we deduce the corresponding equivalent potential components that reproduce the wave equation with a given complex dielectric function. The potential turns out to be energy dependent. For consistency, we assert that we should study the Schrödinger equation with this energy dependent potential in order to coherently compare it with the corresponding wave equation. Actually the effect of inertia in the relativistic K-G equation is by itself a serious issue that distinguishes the behavior of massive particles like electrons in the Schrödinger equation from the wave equation which holds for the massless photons. Our approach simply relates the origin of instability and divergence in the stationary states of both the Schrödinger and wave equations to the time dependence factor $e^{-\mathrm{i}Et/\hbar}$ and $e^{-\mathrm{i}\omega t}$, respectively. Thus we study the pole structure of the transmission in the complex energy or frequency plane and tune the value of the gain (the complex potential) till we see one of the energy eigenvalues in the lower half of the complex energy plane approaches the real energy line to cross it to the upper half. We propose that this cross-over at the critical value of the gain is the origin of the discrepancy between the stationary and time dependent behavior.
This is further supported by the fact that poles in the upper half complex omega plane correspond to bound discrete eigensolutions which have to be included in the expansion along with the eigenstates of the continuous spectrum. It is exactly the omission of this sum over the discrete eigenstates which causes the apparent paradox mentioned before.

## II. FORMULATION OF THE PROBLEM

The propagation of the electromagnetic waves in a medium free of charges and currents is described by the wave equation

$$\left(\vec{\nabla}^2 - \hat{\mu}\,\hat{\varepsilon}\frac{\partial^2}{\partial t^2}\right)\vec{F}(t,\vec{r}) = 0\,, \tag{1}$$

where, $\hat{\mu}$ is the permeability and $\hat{\varepsilon}$ the permittivity of the medium and $\vec{F}$ stands for the electromagnetic fields, $\vec{E}$ or $\vec{B}$. The relative permeability and permittivity $\mu$ and $\varepsilon$ are defined by $\mu = \hat{\mu}/\mu_0$ and $\varepsilon = \hat{\varepsilon}/\varepsilon_0$, where $\mu_0\varepsilon_0 = 1/c^2$ and $c$ is the speed of light in free space. These two parameters are generally complex, space-, and frequency-dependent corresponding to absorbing or active, non-uniform and dispersive medium. The time-independent wave equation for oscillatory electromagnetic fields of the form $\vec{F}(t,\vec{r}) = \vec{F}_0(\vec{r})\, e^{-\mathrm{i}\omega t}$, becomes

$$\left[\vec{\nabla}^2 + \frac{\omega^2}{c^2}\,\mu(\omega,\vec{r})\,\varepsilon(\omega,\vec{r})\right]\vec{F}_0(\vec{r}) = 0\,, \tag{2}$$

assuming that the permittivity and the permeability are piecewise constant. It has been found [5], and our calculations have verified this finding, that there is a discrepancy between the time-dependent and the frequency-dependent solutions of the wave equation in an active medium. More explicitly, the time evolution of a wave packet passing through an active medium of length *L* shows that the transmitted packet is amplified by a factor $e^{aL}$, where *a* is a positive quantity; on the other hand, the frequency-dependent solution shows amplification only up to a critical value of *L*, while for larger values instead of amplification it exhibits attenuation. In this paper we provide an explanation of this paradoxical behavior of the frequency-dependent solution and we reconcile it with the time-dependent solution. Besides the electromagnetic (em) wave equation (2) we study also the relativistic Klein-Gordon (K-G) equation, which, under certain conditions, is equivalent to Eq. (2); its non-relativistic limit reduces to the Schrödinger equation establishing thus the equivalence of all three equations (under certain correspondences). All three equations exhibit the above mentioned discrepancy between the time- and the frequency-dependent solutions in the presence of gain. The resolution of this discrepancy is the same for the three equations.

The time- and the frequency- dependent free K-G equations are given below

$$\left(\vec{\nabla}^2 - \frac{1}{c^2}\frac{\partial^2}{\partial t^2}\right)\psi(t,\vec{r}) = \left(\frac{mc}{\hbar}\right)^2 \psi(t,\vec{r})\,, \tag{3}$$

$$\left(\vec{\nabla}^2 + \frac{\mathcal{E}^2}{\hbar^2 c^2}\right)\psi_0(\vec{r}) = \left(\frac{mc}{\hbar}\right)^2 \psi_0(\vec{r})\,, \tag{4}$$

where $\psi(t,\vec{r}) = \psi_0(\vec{r}) e^{-\mathrm{i}\mathcal{E}t/\hbar}$, $\mathcal{E}$ is the relativistic energy, and *m* is the mass of the particle. In the massless limit these equations look very similar to those above. We confine our discussion to the case where $\mu$ = 1 and to problems in one dimension corresponding to the propagation of plane waves where Eq. (2) is written as

$$\left[\frac{d^2}{dx^2} + \frac{\omega^2}{c^2}\,\varepsilon(\omega,x)\right]\vec{F}_0(x) = 0\,, \tag{2'}$$

Now the effects of the property of the medium, which is contained in the complex function $\varepsilon$, could be incorporated in the one-dimensional version of Eq. (4) by an equivalent effect which is introduced in the form of a complex potential interaction and as follows [6]

$$\left\{\frac{d^2}{dx^2} + \tfrac{1}{\hbar^2 c^2}\left[\mathcal{E} - V(x)\right]^2\right\}\psi_0(x) = \left(\frac{mc}{\hbar}\right)^2 \psi_0(x)\,, \tag{5}$$

where *V* is the time component of a vector potential whose space component is taken to vanish.

Now, for a given medium configuration specified by $\varepsilon(\omega, x)$ and boundary conditions we choose the complex potential *V*(*x*) in Eq. (5) with *m* = 0 that will give the same electromagnetic wave equation, Eq. (2)’. The potential obtained as such will be our guide in writing the non-relativistic Schrödinger equation that gives the quantum mechanical analogue of the wave propagation equation.

Let’s consider the system shown in Fig. 1 where medium I is free space and the waves are incident from left. The constant parameters $\{\varepsilon', \varepsilon'', V', V''\}$ are real. Equation (2)’ in medium II gives:

$$\left[\frac{d^2}{dx^2}+\frac{\omega^2}{c^2}\left(\varepsilon'-\mathrm{i}\varepsilon''\right)\right]F_{II}(x)=0 \tag{6}$$

Notice that as a result of the general relations of $\mathrm{Re}(\varepsilon)$ being even function of $\mathrm{Re}(\omega)$, while $\mathrm{Im}(\varepsilon)$ being an odd function of $\mathrm{Re}(\omega)$ we have for a dispersionless medium with gain that $\mathrm{Re}(\varepsilon)$ is positive for all frequencies, and $\mathrm{Im}(\varepsilon)$ is negative for positive $\mathrm{Re}(\omega)$ and positive for negative $\mathrm{Re}(\omega)$. For a lossy medium it is the other way around. Now, Eq. (5) with $m$ = 0 and $\mathcal{E}=\hbar\omega$, gives.

$$\left[\frac{d^2}{dx^2}+\frac{\omega^2}{c^2}\left(1-2\frac{1}{\hbar\omega}V+\frac{1}{\hbar^2\omega^2}V^2\right)\right]\psi_{II}(x)=0 \tag{7}$$

Comparing these two equations we obtain

$$\varepsilon'=1-2\frac{1}{\hbar\omega}V'+\frac{1}{\hbar^2\omega^2}\left(V'^2-V''^2\right) \tag{8.I}$$

$$\varepsilon''=2\frac{1}{\hbar\omega}V''-2\frac{1}{\hbar^2\omega^2}V'V'' \tag{8.II}$$

Now if the medium is assumed to be non-dispersive (i.e., the permittivity is independent of the frequency $\omega$) then we conclude that the constant potential $V$ should be proportional to $\omega$. In other words the vector potential in the K-G equation is *energy dependent*. It should be proportional to $\mathcal{E}$. Consequently, we write it as $V\equiv\mathcal{E}v$, where $v$ is a dimensionless parameter. Thus, we can now write our previous Eqs.(8) as

$$\varepsilon'=1-2v'+v'^2-v''^2 \tag{9.I)’}$$

$$\varepsilon''=2v''\left(1-v'\right) \tag{9.II)’}$$

Therfore, $v'$ and $v''$ are determined in terms of the parameters $\varepsilon'$ and $\varepsilon''$ as follows

$$\left(1-v'\right)^2=\frac{1}{2}\left(\varepsilon'+\sqrt{\varepsilon'^2+\varepsilon''^2}\right), \tag{10.I}$$

$$v''=\frac{\varepsilon''/2}{1-v'}, \tag{10.II}$$

where for a propagating medium with gain $\mathrm{Re}(1-v)$ is positive and $\mathrm{Im}(1-v)$ has the same sign as the Re($k$). We use the ansatz that $V=\mathcal{E}\left[v'(\varepsilon',\varepsilon'')+\mathrm{i}v''(\varepsilon',\varepsilon'')\right]$ as a guide for our non-relativistic problem. That is, in our investigation of the wave packet propagation through the quantum mechanically equivalent system we take the potential in the Schrödinger equation as $V=E\left(v'+\mathrm{i}\,v''\right)$, where $E$ is the non-relativistic energy and $\{v', v''\}$ are real potential parameters. In Sec. IV, we will also show that this ansatz is supported numerically.

## III. POLES OF THE REFLECTION COEFFICIENT

In the previous section an equivalence was obtained between two representations of the wave equation. One comes from the electromagnetic wave equation and the other comes from the massless limit of the K-G equation with vector potential that embodies the dielectric property of the medium. A necessary condition for the equivalence is that the potential in the K-G equation should be energy dependent. For a dispersionless electromagnetic medium, it had to be proportional to the energy. On the other hand, the permittivity could assume complex values and so could the potential. The nonrelativistic mechanical representation of the system is described by the Schrödinger equation with the same energy-dependent potential. In this section, we study the solution of this Schrödinger equation and calculate the resonance energies by locating the poles of the reflected or transmitted amplitude.

The problem under study is depicted in Fig.1 where a wave packet is incident from left with partial amplitude normalized to unity. Medium I is free space and medium II is a uniform dielectric material whose properties are represented by the uniform potential $V$ in the following one-dimensional time-independent Schrödinger equation

$$\left[\frac{d^2}{dx^2}+\tfrac{2m}{\hbar^2}\left(E-V\right)\right]\psi(x)=0 \tag{11}$$

where $m$ is the inertial mass associated with the wave packet and $E$ is the nonrelativistic energy. The potential is energy-dependent and is written as $V=Ev$, where $v$ is a dimensionless parameter which is generally complex. Note that in the K-G picture we used $\mathcal{E}=\hbar\omega$ as equivalence between the relativistic energy in the massless K-G equation and the frequency in the electromagnetic wave equation. However, this relation no longer holds in the case of the Schrödinger equation representing the motion of an inertial wave packet under the influence of the potential $Ev$. We define the momentum parameter $k$ in relation to the energy $E$ in the usual way as $E=\frac{\hbar^2}{2m}k^2$. It has the dimension of inverse length. Finally, Eq. (11) is rewritten as

$$\left[\frac{d^2}{dx^2}+k^2\left(1-v\right)\right]\psi(x)=0 \tag{12}$$

Its solution for the configuration shown in Fig. 1 is

$$\psi_k(x)=\begin{cases} e^{\mathrm{i}kx}+Re^{-\mathrm{i}kx} & x<0 \\ Ae^{\mathrm{i}k\sqrt{1-v}\,x}+Be^{-\mathrm{i}k\sqrt{1-v}\,x} & 0<x<L \\ Te^{\mathrm{i}kx} & x>L \end{cases} \tag{13}$$

If the incident wave packet $\psi(x\to-\infty)$ is constructed as usual from the partial amplitudes $e^{\mathrm{i}kx}$ using the Fourier expansion $\frac{1}{2\pi}\int_{-\infty}^{+\infty}f(k)e^{\mathrm{i}kx}dk$ for a given choice of momentum distribution $f(k)$, the results are not the same as the solution of the time-dependent Schrödinger equation. Anyway matching the wave function and its gradient at the left boundary ($x$ = 0) and right boundary ($x$ = $L$) we find an expression for the reflection amplitude $R$. It could be written as

$$R=v\frac{e^{2\mathrm{i}\eta\sqrt{1-v}}-1}{\alpha_-^2e^{2\mathrm{i}\eta\sqrt{1-v}}-\alpha_+^2} \tag{14.I}$$

where $\alpha_\pm=\sqrt{1-v}\pm1$ and the dimensionless momentum parameter $\eta$ is defined as $\eta=kL$. If one considers the wave equation Eq. (6) and solves along the same lines for the reflection coefficient the result will be

$$R=\left(\varepsilon-1\right)\frac{e^{2\mathrm{i}\lambda\sqrt{\varepsilon}}-1}{\beta_+^2-\beta_-^2 e^{2\mathrm{i}\lambda\sqrt{\varepsilon}}}, \tag{14.II}$$

where $\beta_\pm=\sqrt{\varepsilon}\pm 1$ and the dimensionless parameter $\lambda$ is defined by $\lambda=\frac{L}{c}\omega$. The similarity between the Schrödinger equation and wave equation results (14) is very apparent under the substitution $1-v\rightarrow\varepsilon$. Thus the parallelism between the wave equation and the corresponding Schrödinger equation is well established in our formalism. Consequently all results that will follow apply equally well to both systems under the prescribed transformation between potential and dielectric constants.

It is, of course, trivial to note from Eq. (14) that the absence of the potential ($v$ = 0, $\varepsilon$ = 1) results in zero reflection. However, interesting are the conditions under which the reflection becomes infinite. These occur when the denominator of Eqs. (14.I) and (14.II) become zero i.e.

$$\left(\alpha_+/\alpha_-\right)^2=e^{2\mathrm{i}\eta\sqrt{1-v}}. \tag{15}$$

The zeros of $R$, other than $v$ = 0, are also interesting but will not be pursued here. When R becomes infinite so does T and there is possibility of having a solution of Eq.(15) without the incident wave, if the Im(k) is positive.
We shall see that this can be realized if there is gain, while it is impossible for a lossy system. Therefore, we take the momentum parameter $\eta$ to be complex and write it as $\eta=\eta'+\mathrm{i}\eta''$. Additionally, we write $\sqrt{1-v}=\gamma'+\mathrm{i}\gamma''$ and

$$\frac{\alpha_+}{\alpha_-}=\frac{\gamma'+1+\mathrm{i}\gamma''}{\gamma'-1+\mathrm{i}\gamma''}\equiv\chi e^{\mathrm{i}\phi}, \tag{16}$$

where for a propagating medium with gain, $\gamma'$ is positive and $\gamma''$ has opposite sign than that of Re(k). We obtain the following solution for (15)

$$\eta''=-\tfrac{\gamma''}{\gamma'}\eta'-\tfrac{1}{\gamma'}\ln\chi \tag{17.I}$$

$$\eta'=\frac{\phi+n\pi-\frac{\gamma''}{\gamma'}\ln\chi}{\gamma'\left[1+(\gamma''/\gamma')^2\right]} \tag{17.II}$$

where $n=0,\pm1,\pm2,...$ which comes from $2n\pi$ difference in the phase of the two complex numbers on both sides of Eq.(15). Therefore, the momentum poles of the reflection amplitude is indexed by the integer $n$ as $k_n=k_n'+\mathrm{i}k_n''$, where $k_n=\eta_n/L$. For the Schrödinger case a wave coming from the left has Re(k) positive. Im(k) is negative for a resonance state and positive for a true discrete bound eigenstate. Thus for a resonance state Im(E) is negative leading to a decaying behavior with time. On the other hand, for a true bound eigenstate Im(E) is positive resulting in a growing behavior with time; it is this growing with time for discrete bound eigenstates which provide the amplification of the wave packet beyond the critical value of L at which the transmission starts decreasing and thus resolves the discrepancy between the time and frequency approaches in systems with gain. Notice that for an electromagnetic wave coming from left, Re(k) can be positive and negative, in contrast to the Schrödinger case. This difference appears because the electromagnetic wave equation is second order in time, while the Schrödinger equation is first order.

## IV. NUMERICAL RESULTS AND DISCUSSIONS

In all our numerical results with Schrödinger equation we use the units for which $\hbar = 2m = 1$ and corresponds to an energy unit $E$ = 0.658 eV, unit of length $L$ = 2.406 $\overset{0}{A}$ and a time unit corresponding to T = $10^{-15}$ s. In Fig.2 we show the numerical results for the poles of the transmission coefficient in the complex $\omega$-plane as obtained from the stationary electromagnetic wave equation for a system of length $L$ and $\varepsilon = 4 - 0.2i$ which is valid for Re(k) positive. This figure compares quite well with Fig.3 which corresponds to the stationary Schrödinger equation with a linearly energy dependent potential. Even though we have presented, in the previous section, a lucid argument that supports our assertion of linear energy dependence of the potential in the Schrödinger equation we have also checked numerically for few other energy dependences that only the linear energy dependence of the potential makes the poles structure of the electromagnetic wave equation similar to the Schrödinger equation. This numerical result strengthens our previous assertion that the Schrödinger equation problem will map onto the wave equation problem only if the potential is linearly dependent on the energy. In Fig.4 we plot the transmission versus the imaginary part of the potential to detect the critical value of gain at which the transmission is maximum for a given system length of 200 in our units and incident energy of 1.209. From this figure we see that the resonance occurs at v" $\approx$ 0.023. In Fig.5 we show the pole that has a real part equal ( or close to ) the incident energy of the wave packet, in our case $E$ = 1.209 ( in our energy units ), and study the behavior of this pole as we increase the imaginary part of the potential in the complex E-plane as shown in Fig.5. It is clear from Fig.5 that, in accordance with our expectations, this important energy pole does cross the real axis at the critical value of the imaginary potential estimated to be v" $\approx$ 0.023 above which the energy pole moves into the upper half of the complex E-plane and consequently causes amplification with time. To support this interpretation we perform the numerical computation of the transmission coefficient from the time dependent Schrödinger equation for values of the imaginary potential below, above and at the critical value as shown in Fig.6. The time dependent Schrödinger equation is being derived from the stationary equation with a linearly dependent potential by substituting E by its operator form $i\hbar \frac{\partial}{\partial t}$ which gives

$$\left[-\frac{\hbar^2}{2m}\frac{\partial^2}{\partial x^2} + i\hbar v \frac{\partial}{\partial t}\right]\psi(x,t) = i\hbar \frac{\partial \psi(x,t)}{\partial t} \qquad (18)$$

The initial waveform used in generating Fig.6 was a Gaussian wave packet of the form

$$\psi(x,0) = e^{-\frac{(x-x_0)^2}{4\sigma^2}} e^{ik_0 x} \qquad (19)$$

centered at $x_0$ with an average momentum of $k_0$ in our units, the normalization constant in (19) does not affect our numerical computations. In all cases we used $\sigma$ = 40.0 units of length. The transmission coefficient was calculated by

$$|T(t)|^2 = \int_L^\infty |\psi(x,t)|^2 \, dx \qquad (20)$$

where $L$ is the system length . It is very clear from Fig.8 that at values of the gain below the critical value the transmission reaches a stationary state at large times while at values of the gain above the critical value the transmission increase steadily but step-like with time leading to an amplified output.

## V. CONCLUDING REMARKS

We believe that our present work sheds light on the real origin of the discrepancy between stationary and time dependent results for quantum or classical systems with gain. While studying the origin of this problem we succeeded in coming up with a mathematical mapping between the Schrödinger equation problem for a massive particle subject to a non hermitian potential and the electromagnetic wave propagation problem in a gain system. The equivalent Schrödinger equation problem resulted in a linearly energy dependent complex potential. All numerical results of different physical quantities supported our ansatz about the similarity between the electronic and photonic systems. Setting this issue aside we have computed the locations of the poles of the reflection and transmission amplitudes in the complex k, or frequency, or energy plane. The poles cross the real axis when the length L ( or, equivalently, the gain ) reaches its critical value. This crossing signals the transfer of the amplification from real k continuous spectrum to the discrete bound eigenstates. Even though our present study is very informative and can be considered as being the first satisfactory explanation of the paradoxial results between time dependent and stationary evolution equation, a complete treatment of wave propagation in gain media can be achieved, in our view, by constructing the time dependent solution from the whole spectrum of the time independent solutions. Thus both the discrete and continuous spectra should be included in the analysis similarly to the approach of Hammer et al. [ 8 ] in dealing with the general solution of Schrödinger equation. The general solution in this case reads :

$$\Psi(x,t) = \int_{-\infty}^{+\infty} \frac{d\omega}{2\pi} f(\omega) g(x,\omega) e^{i\omega t} + \sum_n c_n g_n(x,\omega_n) e^{i\omega_n t} = \oint \frac{dz}{2\pi} f(z) g(x,z) e^{izt} \qquad (21)$$

Where $f(\omega)$ and $c_n$ coefficients are determined from the initial value of $\Psi(x,0) = h(x)$. In the last integral one has to choose a complex contour that will reflect the physical situation at hand. Keeping in mind that $g(x,\omega)$ and $g_n(x,\omega_n)$, solutions of the stationary equation, are not orthonormal to each other because the operator is not hermitian due to the complex nature of $\varepsilon$, we believe that it will not be an easy task to determine analytically or numerically $f(\omega)$ and $c_n$ given h(z).

The numerical results presented in this work constitute an important intermediate step towards a complete resolution of this interesting problem. Notice that the analytical determination of the poles in the transmission ( or reflection ) amplitudes, given by Eq. (17), was greatly facilitated as a result of assuming a dispersionless permittivity. When the dielectric constant become frequency dependent, and this is the real physical situation, since one is dealing with lasing materials and their gain response or lasing action is certainly limited to a finite frequency range hence, $\varepsilon(\omega) = \varepsilon'(\omega) + i\varepsilon''(\omega)$. One possible causal form of this dielectric constant is given by [9]

$$\varepsilon(\omega) = \varepsilon_0 + \frac{f_0}{\omega^2{}_0 - \omega^2 + i\Gamma\omega} \qquad (22)$$

Where $f_0$ is usually small compared to $\varepsilon_0$, $\omega_0$ is the laser resonant frequency and $\Gamma$ represents the width of the response and is related to the lifetime of the lasing mode. All these parameters can be fixed experimentally for a given lasing system. We are planning, in the near future, to implement numerically this dispersive case and trace the path of the main lasing pole in the complex frequency plane.

We conclude by pointing out that, while the time independent em and Schrödinger equations are equivalent under proper correspondences, the same is not true for their time dependent counterparts. Schrödinger equation is first order in time, while the em equation is second order in time. As a result of this the Schrödinger wave packet increases its width linearly with time as it propagates in free space, while the em wave packet retains its shape. This important difference implies that Schrödinger wave packet, in contrast to the em case, never leaves the gain area completely and, hence being amplified without limit as shown in Fig.6.

## ACKNOWLEDGMENTS

The first three Authors acknowledge the support of the physics department at King Fahd University of Petroleum & Minerals. E. N. Economou acknowledges support by the CEC through the IP Molecular Imaging. Fruitful discussions with M. S. Abdelmonem S. A. Ramakrishna and M. Al-Sunaidi are highly appreciated.

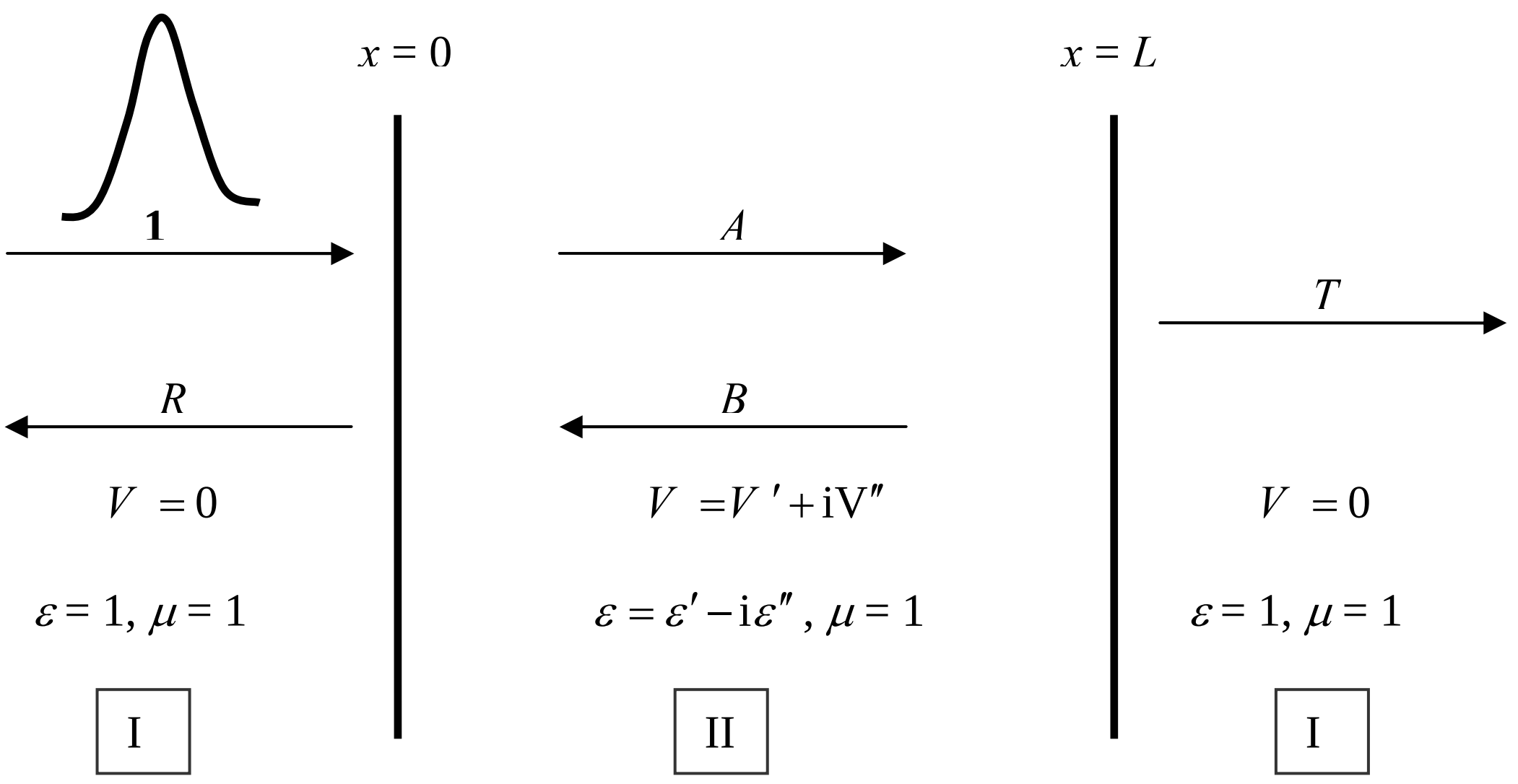

**Fig. 1**

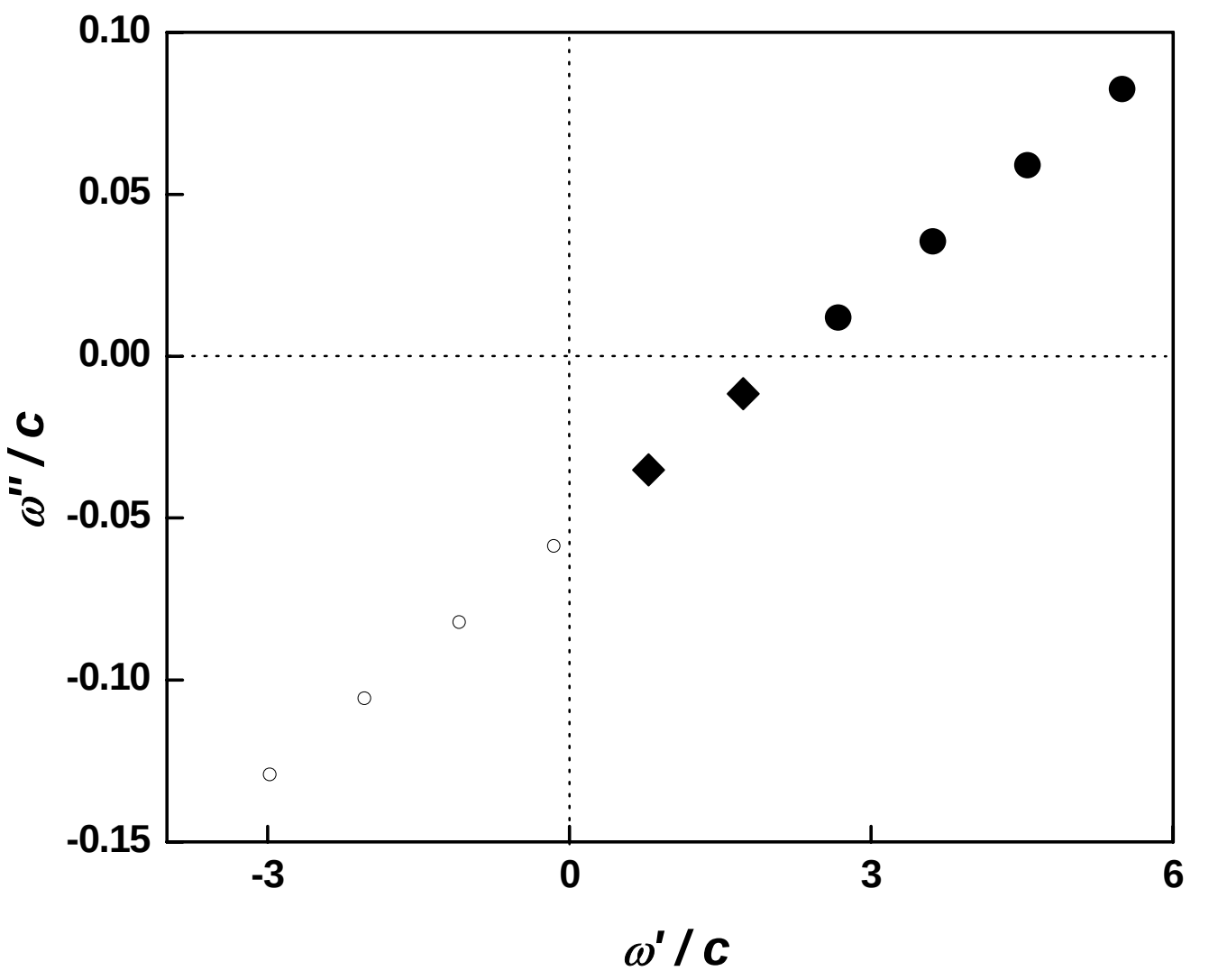

**Fig. 2**: The resonance poles ( diamonds) and the discrete bound eigenstate poles ( filled dots ) of the transmittance obtained from the time-independent classical wave equation in the complex $\omega$-plane with $L$ = 10 and $\varepsilon = 4 - 0.2i$.

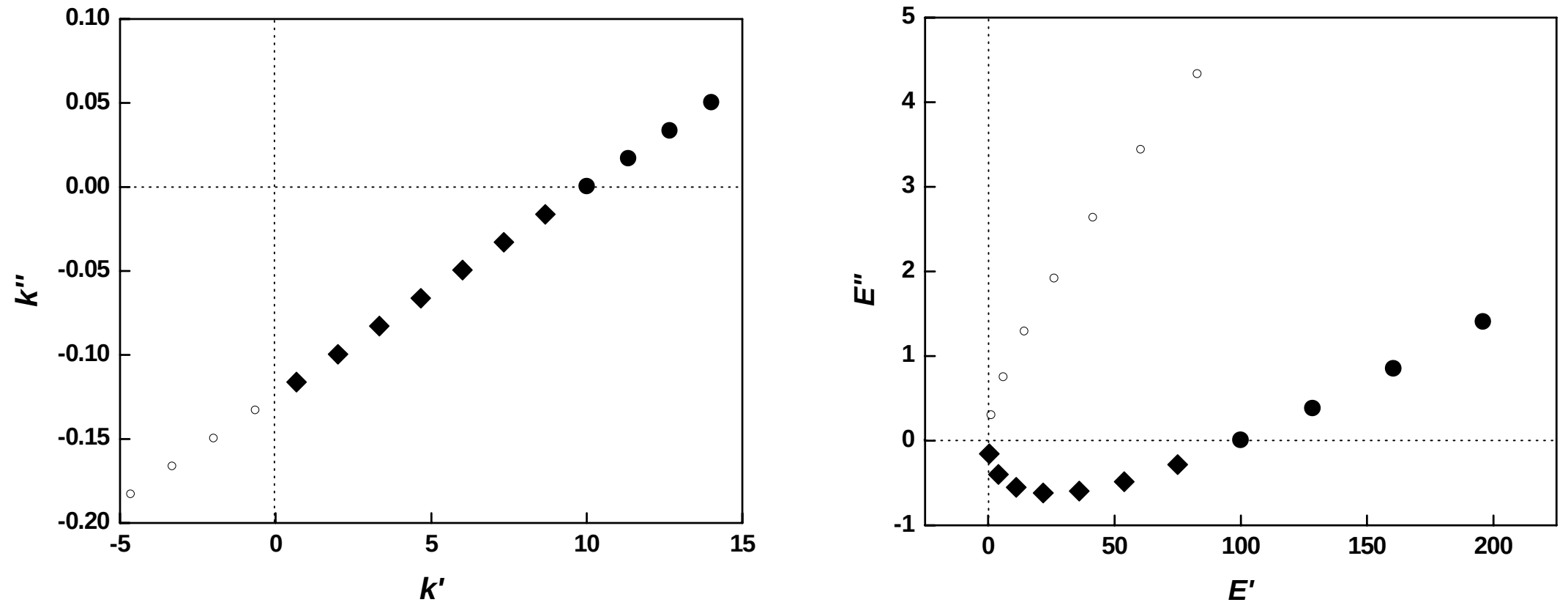

**Fig. 3**: The resonance poles ( diamonds ) and the discrete bound eigenstate poles ( filled circles ) of the transmittance obtained from the time-independent Schrödinger equation in the complex *k*- (left) and *E*- (right) planes for $L$ = 10 and for a potential of the from $V = (-1 + 0.05i)E$ .

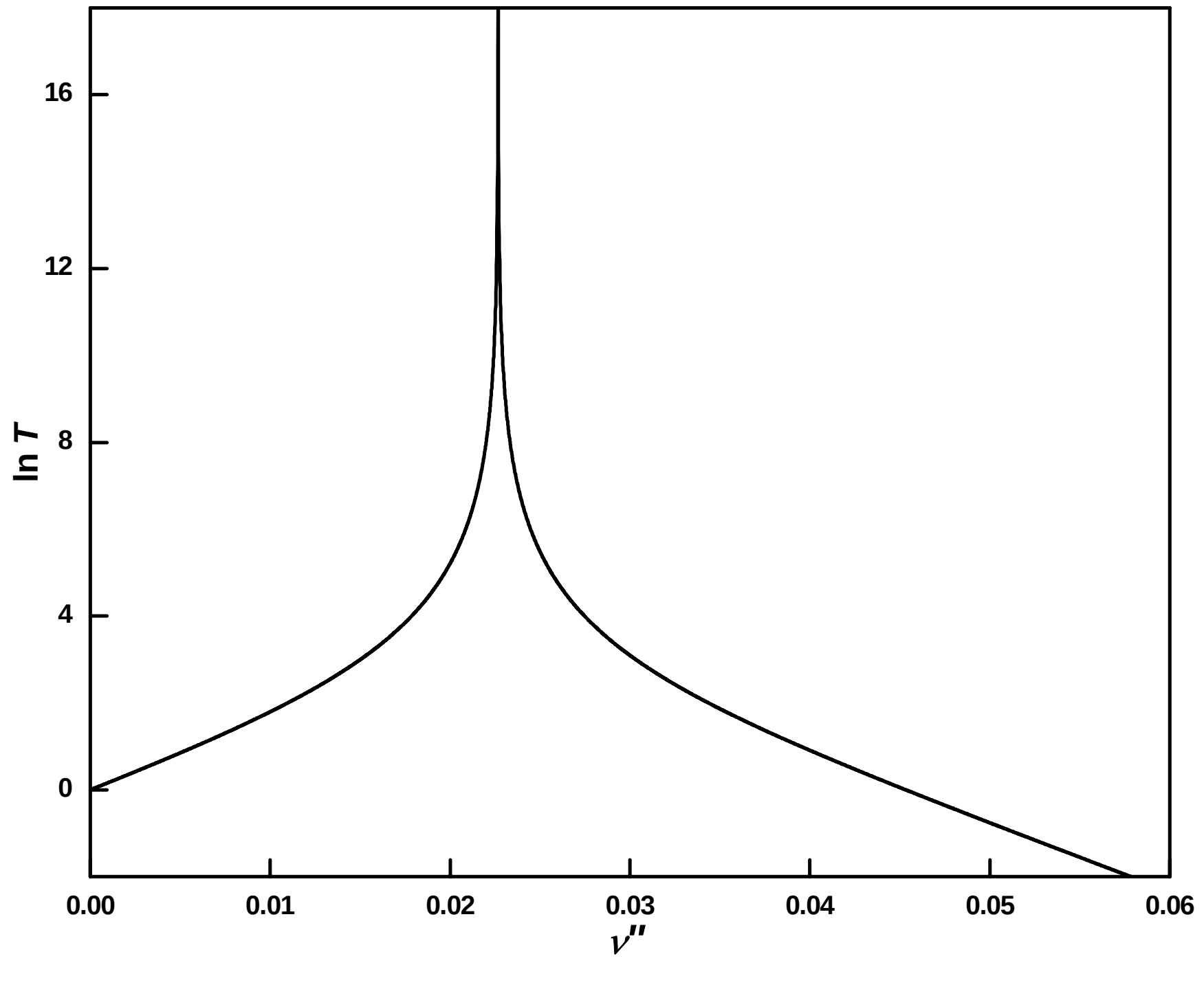

**Fig. 4**: The transmission coefficient obtained from the time-independent Schrödinger equation vs. $v''$ for $E$ = 1.209, $L$ = 200 and $v = -1 + i\ v''$ . The figure shows an extremely sharp peak at $v''_c$ = 0.02267.

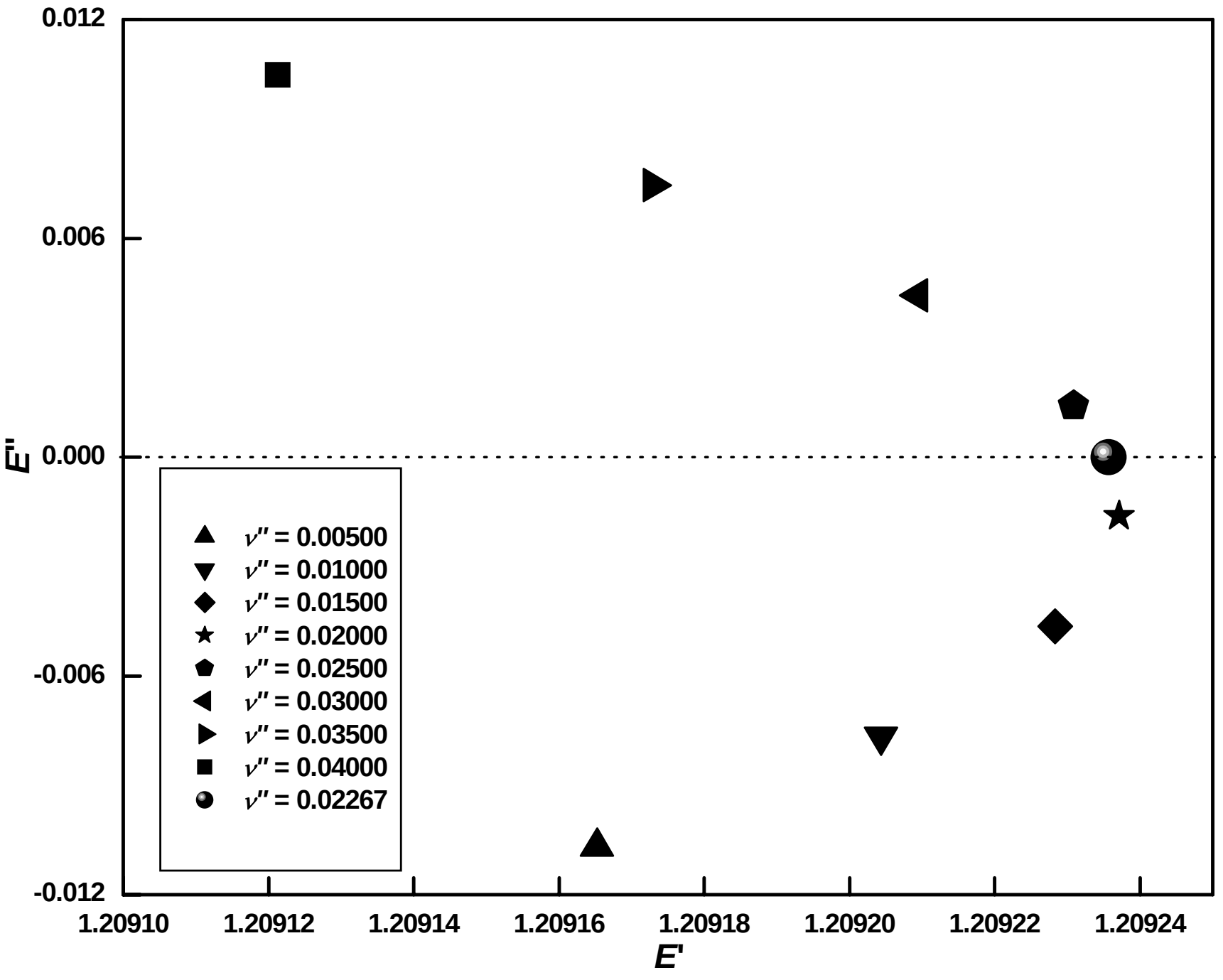

**Fig. 5**: The resonance pole location in the complex $E$ -plane, $L$ = 200 and $v = -1 + i\ v''$ as $v''$ increases. The pole crosses the real axis when $v''_c$ = 0.02267.

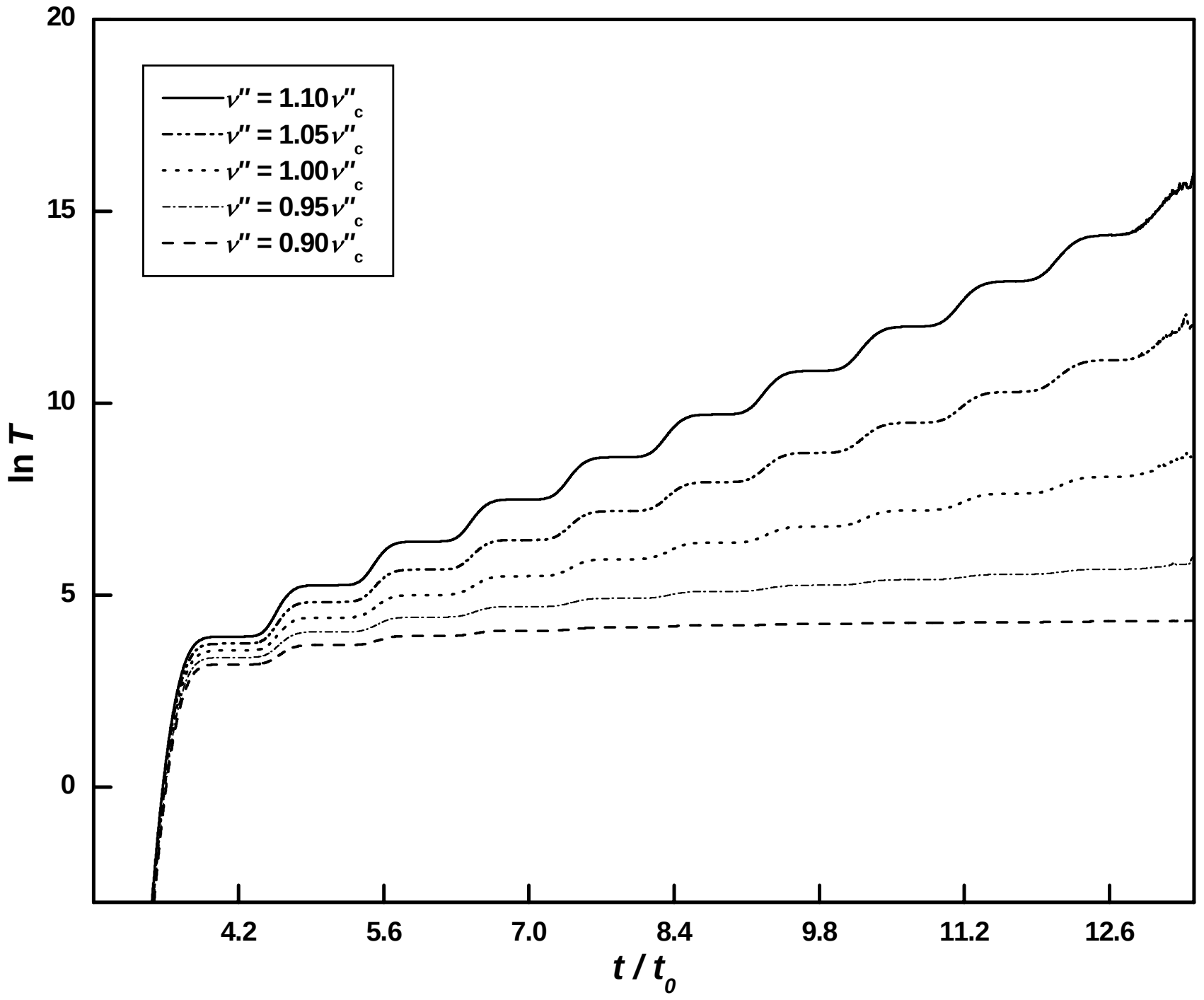

**Fig. 6**: The transmission coefficient obtained from the time-dependent Schrödinger equation for five different values of $v''$ vs. time with $E$ = 1.209, $L$ = 200 and $V = (-1 + i\ v'')E$ , where $t_0 = 143$ time unit.